\documentstyle[psfig,preprint,aps]{revtex}

\begin{document}
\draft

\preprint{ \fbox{NTGMI-95-4 preprint} }

\title{\vspace{.4cm} Rho meson properties in the Nambu-Jona-Lasinio model}
 
\author{A. Polleri$^{a}$\footnote{Present address: Niels Bohr Institute, 
Blegdamsvej 17, 2100 Copenhagen \O, Denmark.}, 
R.A. Broglia$^{a,b,c}$, P.M. Pizzochero$^{a,b}$ 
     and N.N. Scoccola$^{b,d}$ \footnote{Fellow of the CONICET, Buenos
     Aires, Argentina.}}

\address{
     \vspace{.3cm} 
     {\setlength{\baselineskip}{18pt}
     $^a$ Dipartimento di Fisica, Universit\`a degli Studi di Milano,\\
     Via Celoria 16, 20133 Milano, Italy.\\
     $^b$ INFN, Sezione di Milano, Via Celoria 16, 20133 Milano, Italy.\\
     $^c$ The Niels Bohr Institute, University of Copenhagen,\\
     Blegdamsvej 17, 2100 Copenhagen \O, Denmark.\\
     $^d$ Departamento de F{\'{\i}}sica, Comisi\'on Nac. de Energ{\'{\i}}a
     At\'omica,\\ Av. Libertador 8250, 1429 Buenos Aires, Argentina.\\} }

\date{November 1996}

\maketitle

\begin{abstract}
\noindent
Some properties of the rho vector meson are calculated within the
Nambu-Jona-Lasinio model, including processes that 
go beyond the random phase approximation. To classify the higher
order contributions, we adopt $1/N_c$ as expansion parameter.
In particular, we evaluate the leading order contributions to 
the $\rho \rightarrow \pi \pi$ decay width, obtaining the value 
$\Gamma = 118$ MeV, and to the shift of 
the rho mass which turns out to be lowered by 64 MeV with respect to 
its RPA value. A set of model parameters is determined accordingly.
\end{abstract}


\newpage


\section{Introduction}


Vector mesons are known to play an important role in different aspects of
low-energy hadronic physics. In fact, effective lagrangians including nucleons,
pions and vector mesons as fundamental degrees of freedom have been 
successful in describing a large amount of empirical data \cite{Sak69}. On the
other hand, with the introduction of QCD as the fundamental theory of strong
interactions, it became clear that the structure and properties of both baryons
and mesons should be understood in terms of quark and gluon degrees of freedom.
Although much effort has been made in trying to predict low
energy hadron observables directly from QCD, one is still far from
reaching this goal. In such a situation it proves convenient to turn to
the study of effective models. 

One of the models that has received considerable attention in recent years is
the Nambu and Jona-Lasinio (NJL) model \cite{NJL61}. It was originally
proposed as a way to explain the origin of the nucleon mass together with the
existence of the (almost) massless pion in terms of the spontaneous
breakdown of chiral symmetry. Later on, as it became clear that quarks are the
fundamental components of the hadrons, the model was reformulated by replacing
the nucleon fields by quark fields. In its modern version
\cite{VW91,Kle92,HK94,Bij95}, the NJL model is constructed in such a way that
all the symmetries (or approximate low-energy symmetries, such as chiral
symmetry) of QCD are respected. The basic difference with QCD is that gluons
are eliminated in favor of an effective local quark-quark interaction. For
convenient values of the coupling constants, chiral symmetry is spontaneously
broken, a dynamical, ``constituent" quark mass is generated and the pion
appears as a massless Goldstone boson. The pion mass can be brought
to its physical value by adding a small bare, ``current'' quark mass in
the effective lagrangian. 

Due to the local character of the interaction, the
NJL model in $3+1$ dimensions is not renormalizable. Therefore, some kind of
regularization scheme has to be introduced. One of the methods widely used is
to set a cut-off $\Lambda$ in the quark 4-momentum. Typical values are $\Lambda
\approx 1 $ GeV. In a sense, the cut-off gives an idea of the range of
applicability of the model. Since the pion mass is well below $1$ GeV, one might
think that the predictions of the model should be reasonable in that sector. On
the other hand, the rho mass is not far from the typical cut-off values and
one might worry about the validity of the model to describe the properties
of this meson. Within the NJL model, meson modes are
viewed as collective solutions of a Bethe-Salpeter equation in the random
phase approximation (RPA). While the pion can be considered
as a collective mode, the situation concerning the vector mesons is less
clear. Since at the mean field level the dynamical quark mass is
momentum independent, the NJL model does not confine.
Therefore, depending on the model parameters,
the rho mass can be above the unphysical $q \bar q$ threshold and 
consequently its decay into a $q\bar q$ configuration
(Landau damping) is possible within the model. In spite of 
these potential inconveniences, at least at the RPA level, the vector mesons 
seem to emerge as poles in the corresponding ${\cal T}$-matrix channel with 
predicted weak decay constants in  reasonable agreement with their empirical 
values \cite{BM88,KLVW90}. 

To make more definite statements about the validity of the
NJL model in the vector meson sector, one should study other properties. 
In particular the width of the rho meson in its main decay channel, namely
$\Gamma_{\rho \rightarrow \pi\pi}$. This requires, however, to go beyond the RPA
approximation and brings us into the problem of introducing a reasonable 
classification of higher order diagrams. Recently\cite{BHS88,QK94}, it has been
suggested that a convenient way to do so is in terms of an expansion in the
inverse of the number of colors, $N_c$. 
Within QCD, the idea of using $1/N_c$ as an expansion parameter has a long
history. In Refs.\cite{Tho76,Wit79} it was shown that, in order to
have a sensible theory for large values of $N_c$, the quark-gluon coupling 
constant has to scale as $1/\sqrt{N_c}$. It follows that, for example, the 
baryon masses turn out to be of ${\cal O}(N_c)$, the meson masses of ${\cal O}
(1)$, and the coupling of a meson to two-meson states or to $q \bar q$ states of
${\cal O}(1/\sqrt{N_c})$. Independent of these developments, around
the same time very similar ideas
were introduced in a rather different context. Namely, it was shown that the
traditional Hartree + RPA treatment of many-body problems could be considered
as the leading order of an expansion in terms of the inverse of the degeneracy 
of the available Hilbert space, $N$ \cite{BBDLS76}. Moreover, a method to go 
beyond the leading terms based on the $1/N$ expansion was developed 
\cite{BBDLP76}. Since the methods applied in the study the NJL model are 
basically the same as those used in many-body physics, it is quite natural to 
expect that the equivalent of the $1/N$ expansion could be performed. 
In fact, the results of Ref.\cite{QK94} seem to confirm this expectation. 
Some previous estimates of the $\rho$ decay width into two pions within 
the NJL model already exist\cite{BE93KR91}. 
In general, however, these calculations show that this quantity is 
largely underestimated. As we will see, a consistent treatment in terms of 
$1/N_c$ tends to improve this situation. 

In the present paper we report on a calculation of $1/N_c$ corrections
to the RPA description of the $\rho$ meson within the NJL model. 
The paper is organized as follows. In Sec.~II we briefly review the main
features of the NJL model. Namely, we show how quark self-energy and meson
masses, together with their couplings, are obtained in the Hartree + RPA
approximation and we give the values of the model parameters to be used in 
our numerical calculations. 
In Sec.~III we calculate the leading contribution to the $\rho
\rightarrow \pi\pi$ decay width. In Sec.~IV we calculate the $1/N_c$
correction to the rho mass. Conclusions are given in Sec.~V. 
Some details of the   
calculations are collected in Appendices A and B.


\section{Generalized NJL model in the Hartree + RPA approximation}


In this work we consider the two-flavor version (up and down quarks) of
the extended NJL model, defined by the effective Lagrangian 
\begin{equation}
{\cal L}_{NJL} = {\bar \psi}(x) \left( i \rlap/\partial  - m_0 \right) {\psi}(x) + 
{\cal L}_{int}\ .
\label{lnjl}
\end{equation}
In writing Eq.(\ref{lnjl}) we have assumed that the bare  quark
masses are degenerate and set $ m_u = m_d = m_0 $. The four-fermion 
interaction term ${\cal L}_{int}$ is given by 
\begin{eqnarray}
{\cal L}_{int} & = & {G_1\over2} \left[ \left( {\bar \psi}(x) {\psi}(x) 
\right)^2 + \left( {\bar \psi}(x) i \gamma_5 \tau^a {\psi}(x) \right)^2 
\right] \, + \\ 
& & +\, {G_2\over2} \left[ \left( {\bar \psi}(x) \gamma_\mu \tau^a {\psi}(x) 
\right)^2 + \left( {\bar \psi}(x) \gamma_\mu \gamma_5 \tau^a {\psi}(x) 
\right)^2 \right] \nonumber .
\end{eqnarray}
Following the usual treatment, we first apply the mean field formalism in the
Hartree approximation. This leads to the so-called ``gap equation" for the
quark self-energy (also called constituent quark mass) 
\begin{equation}
m_q = m_0 + m_q\,G_1\,N_c\,8i\int^\Lambda\!\!{d^4p\over{{(2\pi)}^4}} {1\over{p^2-m_q^2}}\ .
\label{gapeq}
\end{equation}
The integral in Eq.(\ref{gapeq}) is logarithmically
divergent. Therefore a cut-off $\Lambda$ has to be introduced in order to
regularize it. In the case of the covariant regularization method, that
will be used throughout this paper, this integral has
an explicit analytical form \cite{KLVW90}. 

As well-known, when $m_0=0$ there is a certain critical value of $G_1$ above
which Eq.(\ref{gapeq}) has a non-trivial solution with $m_q\ne0$. This
corresponds to the dynamical breakdown of chiral symmetry. Correspondingly, a
non-vanishing value of the quark condensate $<{\bar \psi} \psi>$ 
develops. For small values of $m_0$, the transition is no longer sharp but 
the same qualitative behaviour is obtained. 

  From Eq.(\ref{gapeq}), it is seen that the dynamically generated 
contribution to $m_q$ in the
Hartree approximation scales with $G_1 N_c$. Therefore, in order to have a
sensible large $N_c$ limit we should impose $G_1 \sim {\cal O}(1/N_c)$. 
This type of requirement is completely equivalent to the one used
in large-$N_c$ QCD. In fact, the scaling $G_1 \sim {\cal O}(1/N_c)$ was to be
expected since the effective quark-quark interaction can be thought of as the
heavy-gluon (i.e. strongly dressed) limit of a gluon mediated interaction.
Therefore, in the Hartree approximation $m_q \sim {\cal O}(1)$. It is easy to
verify\cite{QK94} that the Fock contribution to the  self-energy is of ${\cal
O}(1/N_c)$. 

Making use of the vacuum in the mean field approximation, we shall
now study the associated meson fluctuations. For that purpose, we solve the
Bethe-Salpeter equation for the $\cal T$-matrix
\begin{eqnarray}
i \mbox{\large ${\cal T}$} (q^{2}) & = &\ i \mbox{\large ${\cal K}$} 
 - \ Tr\! \int\!\!{d^4p\over{{(2\pi)}^4}} \ i \mbox{\large ${\cal K}$}
\ i S(p + \mbox{\scriptsize $\frac{1}{2}$} q) \, \times \\
& & i \mbox{\large${\cal T}$} (q^{2})\ i S(p - \mbox{\scriptsize 
$\frac{1}{2}$} q)\, , \nonumber
\end{eqnarray}
where $S(p) = {1\over{\rlap/p - m}}$ is the fermion Feynman propagator. The
$\cal T$-matrix as well as the kernel $\cal K$, can be
decomposed in terms of scalar, pseudoscalar, vector and axial vector
channels.\footnote{In what follows we will concentrate only on the 
pseudoscalar and vector channels, since the others are not relevant 
for our purposes.}

Due to the presence of the explicit chiral symmetry breaking induced by the
current mass $m_0$, we have to deal with the mixing between pseudoscalar and
longitudinal axial fields, called $\pi-A_1$ mixing. Therefore, in the
pseudoscalar channel one obtains a $2 \times 2$ matrix equation, while in the
vector channel a $1 \times 1$ equation is obtained. The solution of these
equations can be found by summing the corresponding geometrical series. One
finally gets 
\begin{equation}
i \mbox{\large ${\cal T}$}\!\!_\pi(q^2) = \left( i \gamma_5 \tau^a \otimes 
i \gamma_5 \tau^a \right) \ \frac{1}{D(q^2)}\, \times 
\end{equation}
\begin{eqnarray}
\times \, \left( \begin{array}
{cc}G_{1} \left( 1 - G_{2}J_{AA}(q^2) \right) &
G_{1}G_{2}J_{PA}(q^2) \\ G_{1}G_{2}J_{AP}(q^2) & G_{2}
\left( 1 - G_{1}J_{PP}(q^2) \right) \end{array} \right) \nonumber \, ,
\end{eqnarray}
where
\begin{eqnarray}
D(q^2) & = & \left( 1 - G_{1}J_{PP}(q^2) \right)
\left( 1 - G_{2}J_{AA}(q^2)\right)\ -  \nonumber \\ 
& & - \ G_{1}G_{2}J_{PA}(q^2)J_{AP}(q^2) \, , \nonumber
\end{eqnarray}
and
\begin{eqnarray}
i \mbox{\large ${\cal T}$}\!\!_\rho(q^2) & = & 
\left( \gamma_\mu \tau^a \otimes \gamma_\nu \tau^a \right) \, \times\\
& & \frac{G_2}{1 -\ G_2\ J_{VV}
(q^{2})} \left( g^{\mu\nu} - q^\mu q^\nu/ q^2 \right)\nonumber \, .
\end{eqnarray}
In these equations, $J_{\alpha\beta}^\Lambda$ are the polarization functions 
(i.e. quark-antiquark loops) of the corresponding channels. They are divergent
functions, that require regularization. For consistency, the same
regularization method used in the mean field approximation is to 
be used here.
Explicit expressions for these functions can be found in the literature
\cite{KLVW90}. 

\begin{figure}
\centerline{\psfig{figure=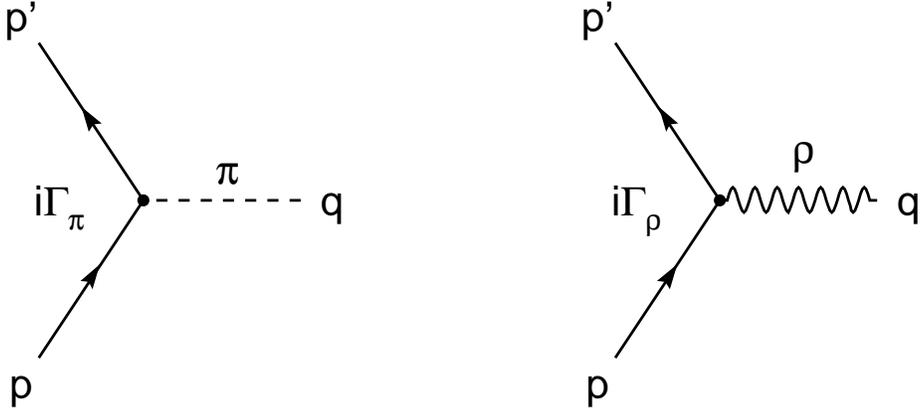,height=7.0cm}}
\protect\caption{ Effective meson-$q$$\bar q$ vertices obtained within the
RPA approximation. We have $i \Gamma_\rho = i g_{\rho - q \bar q}
\gamma_{\mu} \tau^a$ and $i \Gamma_\pi = i g_{\pi - q \bar q} (1 - a_{\pi}
\hat{\rlap/q}) i \gamma^{5} \tau^a$.}
\label{vertex}
\end{figure}

The masses of the bound states and resonances are obtained from the position of
the poles of the $\cal T$-matrix. In order to determine the meson-$q \bar q$
coupling constants, we approximate the $\cal T$-matrix close to the pole
position. We therefore write
\begin{eqnarray}
i\mbox{\large ${\cal T}$}\!\!_{\pi}(q^{2}) & \approx &
\left( i \gamma_5 \tau^a \otimes i \gamma_5 \tau^a \right) \, \times\\
& & ig_{\pi-q \overline q}
(1 + a_{\pi} \hat{\rlap/q})\ \frac{i}{q^{2} - m_{\pi}^{2}}
\ ig_{\pi-q \overline q}(1 - a_{\pi} \hat{\rlap/q})
\nonumber \, ,
\end{eqnarray}
where $\hat{\rlap/q} = \rlap/q /\sqrt{q^2}$, and
\begin{eqnarray}   
i\mbox{\large ${\cal T}$}\!\!_{\rho} (q^{2}) & \approx & 
\left( \gamma_\mu \tau^a \otimes \gamma_\nu \tau^a \right) \, \times\\
& & ig_{\rho-q \overline q}
\ \frac{- i \left( g^{\mu\nu} - q^\mu q^\nu/ q^2 \right)}
{q^{2} - m_{\rho}^{2}}\ ig_{\rho-q \overline q}
\nonumber \, .
\end{eqnarray}
from which the meson-$q$$\bar q$ vertices can be obtained (see
Fig.\ref{vertex}). 

For the vector channel, the rho-quark $g_{\rho-q \overline q}$ coupling 
constant can be defined as the corresponding residue at the pole and expressed
in terms of the function $J_{VV}$. Due to the $\pi-A_1$ mixing, the situation
for the pion is slightly more complicated, but one can still define the 
standard quantities $g_{\pi-q \overline q}$ and $a_{\pi}$ in terms of the 
functions $J_{PA}$ and $J_{AA}$. 

The four parameters of the model can be fixed by fitting four physical 
quantities, namely the quark condensate  $<{\bar q} q>$ 
( $ = \ <{\bar u} u>\ =\ <{\bar d}d>$, since $u$ and $d$ are degenerate), 
the pion and rho 
masses and the pion decay constant $f_{\pi}$. The values we used for the
parameters of the model are 
\begin{tabbing}
-------- \= -------------------------------------- \= ----------- \kill
\                    \> $\Lambda = 1050$ MeV   \> $G_{1}\Lambda^{2} = 10.1$ \\
\                    \> $m_{0} = 3.33$ MeV      \> $G_{2}\Lambda^{2} = -14.4$ 
\end{tabbing}
With this choice, the results in the Hartree + RPA approximation are 
\begin{tabbing}
-------- \= -------------------------------------- \= ----------- \kill
\     \> $<{\bar q} q>^{1/3} \ = -293$ MeV \> $m_q = 463$ MeV \\
\     \> $m_{\pi} = 139$ MeV                \> $g_{\pi-q \overline q} = 4.94$ \\
\     \> $m_{\rho}^{(0)} = 834$ MeV         \> $a_{\pi} = 0.46$ \\
\     \> $f_{\pi} = 93$ MeV              \> $g_{\rho-q \overline q} = 2.12$ 
\end{tabbing}
We point out that the value of the quark condensate, which in the NJL model is 
simply  related to the constituent quark mass $m_q$, is consistent with the
values obtained in lattice gauge calculations. Our prediction for
$m_q$, although large with respect to some recent 
estimates\cite{Bij95,BMO94}, is well within the range of values
usually used in the literature\cite{VW91,Kle92,MRBG94}. With such value of
$m_q$ the RPA prediction for the $\rho$ mass, $m_{\rho}^{(0)}$, 
is below the $q\bar q$
threshold and therefore this particle can be clearly identified
as a pole in the corresponding $\cal T$-channel. Note that $m_{\rho}^{(0)}$ 
is larger than the experimental value $m_{\rho}^{\rm Emp} = 770 $ MeV. As
shown in section~IV, 
the $\rho$ mass will be lowered to this value by self-energy corrections
\footnote{Higher order corrections might, of course, also affect the quark 
and pion properties. However, due to the constraints imposed by chiral symmetry
in those channels, the calculation of such corrections is a quite 
delicate matter\cite{DSTL95}. 
Here, we prefer to ignore these effects and 
choose our parameters in such a way that the RPA predictions for the pion 
properties agree well with the empirical values.}.

Before concluding this section it is worth recalling 
the $N_c$-behaviour of the
meson properties obtained in the RPA approximation. It can be shown that all
the polarization functions are proportional to $N_c$. As a consequence, the
meson masses are of ${\cal O}(1)$, while the meson-$q$$\bar q$ coupling
constants are of ${\cal O}(1/\sqrt{N_c})$. This is in 
agreement with the $N_c$
counting rules obtained in QCD \cite{Tho76,Wit79} 
and also with those found in many-body theories \cite{BBDLS76,BBDLP76}. 


\section{The $ \rho \rightarrow \pi \pi$ decay}

\begin{figure}
\centerline{\psfig{figure=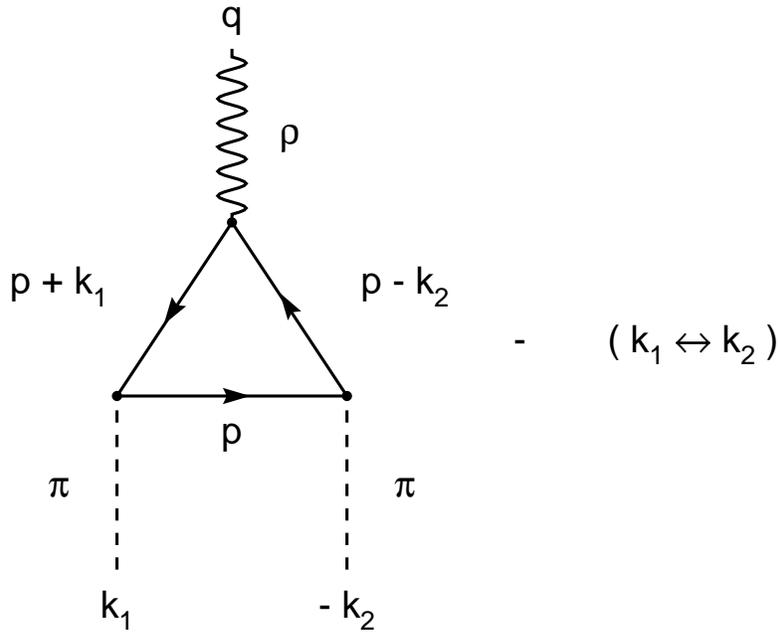,height=10.0cm}}
\protect\caption{ The $\rho - \pi \pi$ amplitude used to calculate the 
decay width. It corresponds to the function $T^{\mu}(q;k_1,k_2)$ as in
Eq.(\protect\ref{ampl}).}
\label{ampgraph}
\end{figure}

The main decay channel of the rho meson is
into two pions. The leading contribution to this process corresponds to the
diagram shown in Fig.\ref{ampgraph},  
which is of ${\cal O}(1/\sqrt{N_c})$ in agreement with the large $N_c$ QCD
result. This behaviour results from the factor $N_c$ due to the quark loop 
and from the factors $1/\sqrt{N_c}$ associated with the 
meson-$q$$\bar q$ vertices (cf. Fig.\ref{vertex}).

Using the usual Feynman rules, one obtains the amplitude for the decay 
process 
\begin{equation}
\mbox{\large ${\cal M}$}(q;k_1,k_2) = \epsilon_{\mu}(q) \
T^{\mu}(q;k_1,k_2)\ ,
\end{equation}
where $q = k_1 + k_2$, and 
\begin{eqnarray}
\label{ampl}
T^{\mu}(q;k_1,k_2) & = &
 -\, tr \!\! \int \!\! {d^4p\over{{(2\pi)}^4}} \, i g_{\rho - q \bar q} 
\gamma_{\mu} \, \frac{i}{\rlap/p - \rlap/k_2 - m_q} \times \\
& & i \sqrt{2} g_{\pi - q \bar q} (1 + a_{\pi} 
\hat{\rlap/k_2}) i \gamma^{5} \frac{i}{\rlap/p  - m_q} \times \nonumber \\
& & i \sqrt{2}
g_{\pi - q \bar q} (1 + a_{\pi} \hat{\rlap/k_1}) i \gamma^{5} \times
\nonumber \\
& & \frac{i}{\rlap/p + \rlap/k_1 - m_q} \  -\  (k_{1} 
\leftrightarrow k_{2}) \nonumber\, .
\end{eqnarray}
Evaluating the trace one gets
\begin{eqnarray}
\label{aux}
T^{\mu}(q;k_1,k_2) & = & -\,2\,i\,N_c\,g_{\rho - q \overline q}
\ g_{\pi - q \bar q}^2 \, \times \\
& & \mbox{\large [} k_1^\mu\ G(q^2;k_1^2,k_2^2) - k_2^\mu\ 
G(q^2;k_2^2,k_1^2) \mbox{\large]}\nonumber \, ,
\end{eqnarray}
where the function $G(q^2;k_1^2,k_2^2)$ can be written in terms of the 
standard regularized functions which appear in the NJL model.
Its explicit expression is given in Appendix A.

Making use of Fermi's Golden Rule, the corresponding
decay width is found to be 
\begin{equation}
\mbox{\large $\Gamma$} = \frac{1}{2 m_{\rho}}\ \int d \phi^{(2)} 
\ \overline{\left| \mbox{\large $\cal M$} \left( m_{\rho}^{2},
m_{\pi}^{2} \right) \right|^{2}}\ ,
\label{decay}
\end{equation}
where $\int \!d \phi^{(2)} = \int \!\frac{d^{3}k_{1}}{(2 \pi)^{3} 2 E_{k_{1}}}
\frac{d^{3}k_{2}}{(2 \pi)^{3} 2 E_{k_{2}}}${\footnotesize $(2 \pi)^{4}
\delta^{4} (q - k_{1} - k_{2})$} is the standard two-body phase-space-measure.

The bar over the squared transition amplitude implies an average over the
initial polarizations and a sum over the final ones. Moreover, all particles 
are put on their mass shells. The calculation of expression (\ref{decay}) 
is straightforward and details are given in Appendix A. 

One obtains
\begin{eqnarray}
\label{pippo}
\label{width}
\mbox{\large $\bf\Gamma$}_{\rho \rightarrow \pi \pi} & = &
\frac{{N_c}^2}{12 \pi}\,g_{\rho - q \overline q}^{2}\ g_{\pi - q \overline 
q}^{4} \ \, m_{\rho} \left( 1 - \frac{4 m_{\pi}^{2}}{m_{\rho}^2} \right)^{3/2}
\times \\
& & G^2 \! \left( m_\rho^2;m_\pi^2 \right) \nonumber \, .
\end{eqnarray}

In order to be fully consistent with the $1/N_c$ expansion, all the
quantities appearing in Eq.(\ref{width}) have to be taken at the values 
obtained in Hartree + RPA approximation, namely those given at the end of 
the previous section. In particular, we have to take RPA value for
the $\rho$ mass. This is equivalent 
to use the Rayleigh-Schr\"oringer (RS) perturbation theory \cite{Dal61}. 
The final result is
\begin{equation}
\mbox{\large $\bf\Gamma$}_{\rho \rightarrow \pi \pi}^{\rm Th} =
\ 118 \ {\rm MeV .}
\end{equation}
As we see, already at leading order in $1/N_c$ we obtain a value
which compares reasonably well with empirical value \cite{PDT94}
\begin{equation}
\mbox{\large $\bf\Gamma$}_{\rho \rightarrow \pi \pi}^{\rm Emp} =
\ 151.2 \pm 1.2 \ {\rm MeV .} 
\end{equation}

It is interesting to note that, instead of using the RPA value,
we could have evaluated the width using the corrected
$\rho$ mass (i.e. its value once $1/N_c$ corrections are included;
see next section). This corresponds to use the so-called 
Brillouin-Wigner (BW) perturbation theory\cite{Dal61}. 
This type of expansion implies the inclusion of  
diagrams of order higher than $1/\sqrt{N_c}$, in which a $\rho$ meson line
appears as an intermediate state. At least in the case of many-body 
systems, this leads to a poor convergence of the perturbative
series \cite{BBB76} since   
for a given order in $1/N_c$, such diagrams usually have opposite 
signs with respect to all the other (disregarded) diagrams of the same order.
Within this scheme, the resulting width is
\begin{equation}
\mbox{\large $\bf\Gamma$}_{\rho \rightarrow \pi \pi}^{\rm Th} ({\rm BW}) =
\ 96 \ {\rm MeV ,}
\end{equation}
which is in somewhat worse agreement with the empirical
value.

We conclude this section with a short remark on the evaluation of
$\mbox{\large $\bf\Gamma$}_{\rho \rightarrow \pi \pi}$ 
using the naive low momentum expansion, as sometimes done in the literature.
Within this approximation, the form factor $G^2 \! \left( q^2;m_\pi^2 \right)$ 
in Eq.(\ref{width}) is replaced by its value at $q^2=0$. This is based on the assumption that this function is only weakly dependent on $q$ and leads, of course, to an important simplification in the evaluation of the corresponding decay amplitudes. In order to investigate whether this approximation
could have been used in the present calculation we display in Fig.(3) 
the momentum dependence of $G^2 \! \left(q^2;m_\pi^2 \right)$ as calculated 
using Eq.(\ref{formshell}). As we see, in the range of momenta
we are interested in ( $q \approx 800 \ MeV$), 
$G^2 \! \left( q^2;m_\pi^2 \right)$ differs from 
its value at zero momentum for about $30 - 40 \%$. This clearly indicates 
that, at least within the present regularization scheme, 
the use of the low momentum approximation would have been 
inappropiate.


\section{Self-energy correction to the rho mass}


The leading self-energy corrections (i.e. beyond RPA) to the rho meson
propagator are of order $1 / N_c$, as can be seen from the corresponding 
diagrams displayed in Fig.\ref{bubbles}.
The self-energy, which we will call ${\bf \Pi}^{\mu \nu} (q)$, can 
be decomposed into its real and imaginary parts which can be 
calculated using cutting rules and dispersion relations. 

As already mentioned, the NJL model does not confine, i.e. it allows hadrons 
to decay into free quarks. In order to cure this unphysical behaviour 
we follow Ref.\cite{CSS92} and impose 
confinement ``by hand'', that is we neglect contributions from cuts of the
diagrams across quark lines thus taking into account only cuts across meson 
lines. This is consistent with the pole approximation for the meson 
lines already used in the previous sections. Within such an approximation, the
diagrams (b) and (c) of Fig.2 do not contribute, while only the cut across 
the pion lines in diagram (a) will give a contribution to the imaginary part 
of ${\bf \Pi}^{\mu \nu} (q)$. Therefore
\begin{eqnarray}
\label{polar}
- \mbox{\small $\frac{1}{i}$}
{\bf \Pi}^{\mu \nu}(q) & = & \int \!\! {d^4p\over{{(2\pi)}^4}} 
\, i T^{\mu}(q;p + \mbox{\scriptsize $\frac{1}{2}$} q,p - \mbox{\scriptsize 
$\frac{1}{2}$}q) \, \times\\
& & \frac{i}{(p - \mbox{\scriptsize $\frac{1}{2}$} q)^2 - m_\pi^2}
i T^{\mu}(q;p - \mbox{\scriptsize $\frac{1}{2}$} q,p + 
\mbox{\scriptsize $\frac{1}{2}$} q) \times \nonumber \\
& & \frac{i}{(p + \mbox{\scriptsize $\frac{1}{2}$} q)^2 - m_\pi^2}
\, , \nonumber
\end{eqnarray}
where $ i T^{\mu}$ is the $\rho - \pi \pi$ amplitude of 
Eq.(\ref{ampl}) and it represents the quark triangular loops. Making use of the 
cutting rules, one obtains 
\begin{equation}
{\cal I}m\  {\bf \Pi}^{\mu \nu} (q) = - \frac{1}{2} \int
d \phi^{(2)} \ i T^{\mu}(q) \, \left( i T^{\nu}(q) \right)^{\ast} ,
\label{cutting}
\end{equation}
where the pion momenta $k_1$ and $k_2$ are set on shell. Using the 
transversality  property of the self-energy, that is ${\bf \Pi}^{\mu \nu} (q) = \left( 
g^{\mu\nu} - q^\mu q^\nu/ q^2 \right) {\bf \Sigma} (q^2)$, it is possible to write 
(see Appendix B) 
\begin{equation}
{\cal I}m\  {\bf \Sigma} (q^2) = -\ q \, \mbox{\large $\Gamma$}_\rho(q)\ ,
\label{imagine}
\end{equation}
where $\mbox{\large $\Gamma$}_\rho(q)$ has the same expression as in 
Eq.(\ref{decay}) except that the rho momentum is not on shell. 

The real part of the self-energy can be calculated
from the above expression making use of the Kramers-Kr\"{o}nig relation, 
leading to
\footnote{Strictly speaking eq.(\ref{real}), which becomes
exact in the limit $\Lambda^2 \rightarrow \infty$, defines an
approximation to ${\cal R}e\ {\bf \Sigma} (q^2)$.}
\begin{equation}
{\cal R}e\ {\bf \Sigma} (q^2) = \frac{1}{\pi} {\cal P} 
\int_{4 m_{\pi}^2}^{\Lambda^2} d \mu^2 \, \frac{{\rm Im} \ {\bf \Sigma} (\mu^2)}
{\mu^2 - q^2}\ . 
\label{real}
\label{sel}
\end{equation} 
It should be noticed that {\it a priori} there is no reason to use
in Eq.(\ref{sel}) the same cutoff as in, for example, Eq.(\ref{gapeq}).  
However, if one assumes that the cutoff represents the momentum scale up 
to which the low energy effective model can be used, the fact that
certain particles can be described up to momenta higher than such scale
while others not is quite hard to justify. Consequently, we choose to
{\it define} our model by using the same cutoff for all the possible
types of loops.

The pole position is now moved in the complex plane with respect to the RPA 
value, the imaginary part being related to
the decay width and the real part with the mass of the particle.
Then, the expression of the rho mass which includes corrections up to 
order $1 /N_c$ reads 
\begin{equation}
m_\rho^2 =  (m_{\rho}^{(0)})^2 +  
{\cal R}e \ {\bf \Sigma} \left( (m^{(0)}_\rho)^2 
\right) .  
\label{correc}
\end{equation}
Using Eq.(\ref{correc}) together with our set 
of model parameters, we find a shift of $- 64 {\rm\ MeV}$ from the 
RPA value. The new, ``physical" rho mass is therefore
\begin{equation}
 m_{\rho} = 770 \ {\rm MeV},
\end{equation}
which justifies {\em a posteriori} the choice made at the end of section~II.
Here again we have used RS perturbation theory to evaluate the mass shift.

The use of BW perturbation theory would correspond to solve the
equation
\begin{equation}
m_\rho^2 =  (m_{\rho}^{(0)})^2 +  {\cal R}e \ {\bf \Sigma} \left( m_\rho^2 
\right) .  
\end{equation}
As we have already stressed this method is not
completely consistent with the $1/N_c$ expansion. Numerically,
however, the resulting shift $- 60 {\rm\ MeV}$ is not so different from 
that obtained by using RS perturbation theory. In a way this behaviour
reminds of that found in the study of nuclear giant resonances,
where the predictions for the resonance width are more sensitive 
than the shift of its centroid to the perturbation method used.

The values of the rho mass shift reported above agree well 
with those obtained in ref. \cite{KKW96} using an effective mesonic action.
On the hand, arguments based on the assumption that the $\rho_0-\omega$
mass splitting ( $\approx 12 \ MeV$ empirically) is basically due to the rho
mass shift (see e.g. ref.\cite{MT96}) seem to indicate that our value is too large. It is clear that since the origin of such splitting is not yet fully
understood this type of argument has to be taken with some care. In any case, 
if one relaxes the constraint imposed above on the rho momentum cut-off and
reduces it by about $15 \%$, a new self-consistent calculation shows that the
shift can be
reduced down to a value consistent with the $\rho_0-\omega$ mass splitting. 
This variation has only a minor effect on the predicted rho decay
width which is lowered by about $10 \%$.


\section{Conclusions}   


In this paper we have evaluated the rho decay width into two pions and the 
corresponding mass shift within the Nambu and Jona-Lasinio model.
Since the calculation of these quantities requires to go beyond
the usual Hartree + RPA approximation, we have introduced
the inverse of the number of colors $1/N_c$ as a good expansion 
parameter. In terms of this parameter, the Hartree + RPA approximation 
corresponds to the inclusion of diagrams of ${\cal O} (1)$, while
leading contributions to the decay width and mass shift 
are of ${\cal O} (1/N_c)$. In order to be fully consistent
within $1/N_c$, we have used the values obtained in the Hartree + RPA
approximation to compute the $1/N_c$ quantities.
This is equivalent to the use of Rayleigh-Schr\"odinger perturbation 
theory in many-body physics.

The calculated decay width turns out to be $118$ MeV to be 
compared with the experimental value $151.2 \pm 1.2$ MeV.
This corresponds to a $20 \%$ accuracy, which is what one could 
expect from a leading order calculation. For the mass shift we
have obtained $- 64$ MeV, which is of the order of $10 \%$ of
the RPA value.  
This result, together with those obtained
in the quark sector \cite{QK94}, seems to indicate a rather good 
convergence of the perturbative series in $1/N_c$. 
In a way, this fact also justifies
{\it a posteriori} our choice of model parameters, which was
based on the assumption that next-to-leading corrections
to the mass shift were negligible. Of course, to make more
definite statements about the convergence of the $1/N_c$
expansion other properties as well as higher orders in the
expansion would have to be calculated.

In the evaluation of the mass shift we have
disregarded the diagrams that contain cuts across quark lines. 
Such an approximation, which is consistent with the pole
approximation, is certainly a rather primitive way to implement 
confinement.
Quite recently\cite{BB95}, various extension of the NJL model 
in which quarks are confined through a some effective 
confinement mechanism have been considered. 
An obvious extension of our study would
be the evaluation of the rho meson properties in such type of 
models. In fact, shortly after the present paper was submitted for
publication, some work along this line has been reported\cite{MT96}. 

We conclude that the Nambu-Jona-Lasinio model provides an economic
and at the same time reasonably accurate picture of the $\rho$ meson,
in particular of its decay width into two pions. This is true, provided
that a 
single cut-off energy of the order of 1 GeV is introduced to 
regularize the different ultraviolet divergences and that 
cuts in the diagrams are taken only 
across meson lines, to avoid decay into free quarks. With these 
approximations, all the many-body techniques can be used to systematically
explore the consequences of the variety of couplings of meson among
themselves, as well as with fermions.


\section*{Acknowledgements}

We would like to thank P.F. Bortignon and W. Weise for enlightening
discussions.

\appendix

\section{The $\rho - \pi \pi$ Amplitude and Decay.}


As seen in Sec.III, the $\rho \rightarrow \pi \pi$ decay amplitude 
can be expressed in terms of
\begin{eqnarray}
T^{\mu}(q;k_1,k_2) & = & -\,2\,i\,N_c\,g_{\rho - q \overline q}
\ g_{\pi - q \bar q}^2 \, \times \\
& & \mbox{\Large $[$} k_1^\mu\ G(q^2;k_1^2,k_2^2) - 
k_2^\mu\ G(q^2;k_2^2,k_1^2) \mbox{\Large $]$} \nonumber \, .
\end{eqnarray}
Here
\begin{eqnarray}
\label{form}
G(q^2;k_1^2,k_2^2) = G^{(0)}(q^2;k_1^2,k_2^2) \, +\, h_\pi 
\mbox{\Large $[$} I_2(k_1^2)\ - \hspace{1.5cm} & & \\
- \ 2\,G^{(0)}(q^2;k_1^2,k_2^2) \mbox{\Large $]$} \, +\, h_\pi^2 
\mbox{\Large $[$} - I_2(k_1^2) + G^{(0)}(q^2;k_1^2,k_2^2) \ + & & \nonumber\\
+\ \frac{(q^2 - k_1^2 + k_2^2)}{8\,m_q^2\,q^2} \ J_{VV}(q^2) \, 
\mbox{\Large $]$} \, ,\hspace{4.15cm} \nonumber
\end{eqnarray} 
with $h_\pi = \frac{2m_q}{m_\pi}\,a_\pi$ and
\begin{eqnarray}
\label{formzero}
G^{(0)}(q^2;k_1^2,k_2^2) = I_2(k_1^2) + \frac{1}{q^2 - 2(k_1^2 + k_2^2)}
\, \times \hspace{1cm} & & \\
\mbox{\huge $\{$} \left[ q^2 - (k_1^2 + k_2^2) \right] \mbox{\Large $[$} 
\, I_2(q^2) -\, \left( \frac{I_2(k_1^2) + I_2(k_2^2)}{2} \right) 
\mbox{\Large $]$} \ + & & \nonumber \\
+\ (3 k_2^2 - k_1^2) \, \left( \frac{k_2^2 + k_1^2}{2} \right) 
I_3(q^2;k_1^2,k_2^2) \mbox{\huge $\}$} \, . \hspace{1.6cm} \nonumber
\end{eqnarray}
In the last two equations we have used  
\begin{equation}
I_2(q^2) = 4i \int^\Lambda \!\! {d^4p\over{(2\pi)^2}} \, {1\over{[p^2-m_q^2] 
[(p-q)^2-m_q^2]}} \, , 
\end{equation}
\begin{eqnarray}
I_3(q^2;k_1^2,k_2^2) & = & 4i \int^\Lambda \!\! {d^4p\over{(2\pi)^2}} \,
{1\over{[(p-k_1)^2-m_q^2]}}\, \times \\
& & {1\over{ [p^2-m_q^2] [(p+k_2)^2-m_q^2] }} \nonumber \, , 
\end{eqnarray}
\begin{equation}
J_{VV}(q^2) = {2 N_c\over3} \left[ (q^2 + 2 m_q^2) I_2(q^2) - 2 m_q^2 I_2(0) 
\right] \, ,
\end{equation}
where $q = k_1 + k_2$. When pions are on shell, one obtains a much simpler 
expression 
\begin{equation}
T^{\mu}(q) = -\,2\,i\,N_c\,g_{\rho - q \overline q}
\ g_{\pi - q \bar q}^2\ G(q^2; m_\pi^2) \left( k_1 -  k_2 \right)^\mu \! ,
\label{ampshell}
\end{equation}
where now we have
\begin{eqnarray}
\label{formshell}
G(q^2; m_\pi^2) = ( 1 - h_\pi ) \mbox{\huge $\{$} I_2(m_\pi^2) + 
\frac{1 - h_\pi}{q^2 - 4m_\pi^2}\, \times \hspace{1.3cm} & & \\ 
\left[ (q^2 - 2 m_\pi^2 ) \left( I_2(q^2) - I_2(m_\pi^2) \right) + 
2 m_\pi^4 I_3(q^2;m_\pi^2) \right] \ - \nonumber & & \\
-\ \frac{h_\pi^2}{2\,(1 - h_\pi)}\ \frac{J_{VV}(q^2)}{4\,m_q^2}
\mbox{\huge $\}$} . \hspace{4.4cm} \nonumber
\end{eqnarray}
This pion-on-shell expression agrees with the result obtained in
Ref.\cite{BMO94}. Note, however, the different type of regularization
used in that reference.

To calculate the decay width, one has to evaluate
\begin{eqnarray}
& & \int \! d\phi^{(2)} \ \overline{\left| \mbox{\large $\cal M$}
(m_{\rho}^{2},m_{\pi}^{2})\right|^{2}} \\
& & \hspace{2.5cm}= \int \! d\phi^{(2)} 
\!\! \sum_{\lambda = 0,+,-}\!\!
\left| \,\epsilon^{(\lambda)}_{\mu}(q) \ T^{\mu}(q) \right|^2 , \nonumber
\end{eqnarray}
which can be easily done by using the relations
\begin{equation}
\sum_{\lambda = 0,+,-}\!\! \epsilon^{(\lambda)}_{\mu}(q)
\ \epsilon^{(\lambda)}_{\nu}(q) = - \left( g^{\mu\nu} - q^\mu q^\nu/ q^2
\right)
\end{equation}
and
\begin{equation}
\int \! d\phi^{(2)} = \frac{1}{8 \pi}
\ \sqrt{1 - \frac{4 m_{\pi}^{2}}{m_\rho^2}}\ ,
\label{phase}
\end{equation}
obtaining as final result the expression given in Eq.(\ref{width}).


\section{The rho self-energy}


  From Eq.(\ref{polar}) in the text we have, relabelling momenta,
\begin{eqnarray}
{\bf \Pi}^{\mu \nu}(q) & = & - i \! \int \!\! {d^4k_1\over{{(2\pi)}^4}} 
{d^4k_2\over{{(2\pi)}^4}} \ (2\pi)^4\,\delta^4(q - k_1 - k_2) \, \times \\
& & i T^{\mu}(q;k_1,k_2)\, \frac{i}{k_2^2 - m_\pi^2}
\,i T^{\mu}(q;k_2,k_1)\, \times \nonumber \\
& & \frac{i}{k_1^2 - m_\pi^2}\ . \nonumber
\end{eqnarray}
Cutting the diagram across pion lines, it is possible to obtain the imaginary
part of the self-energy. This consists in re-writing the amplitude with
appropriate rules and putting pions on shell with the prescription
\begin{equation}
\frac{i}{k^2 - m_\pi^2} \longrightarrow \frac{1}{(2\pi)^4} 2\pi\,\delta(k^2 -
m_\pi^2)\,\theta(k^0)\ .
\end{equation}
Using Eq.(\ref{cutting}), we can write
\begin{equation}
{\cal I}m\  {\bf \Sigma} (q^2) = - \frac{1}{6} \int
d \phi^{(2)} \ \left| i T^{\mu}(q) \right|^2\ .
\end{equation}
Making use of Eq.(\ref{phase}) for off shell $\rho$ momentum, it is possible 
to evaluate the integral in the rhight hand side. One obtains
\begin{eqnarray}
{\cal I}m\ {\bf \Sigma} (q^2) & = &
\frac{{N_c}^2}{12 \pi}\,g_{\rho - q \overline q}^{2}\ g_{\pi - q 
\overline q}^{4}\ q^2 \left( 1 - \frac{4 m_{\pi}^{2}}{q^2} \right)^{3/2}\,
\!\!\!\!\!\!\! \times \\
& & G^2(q^2;m_\pi^2) \nonumber
\end{eqnarray}
and therefore the result in Eq.(\ref{imagine}).

                 
                 


\vspace*{2cm}

\vspace*{1cm}

\pagebreak

\begin{figure}
\centerline{\psfig{figure=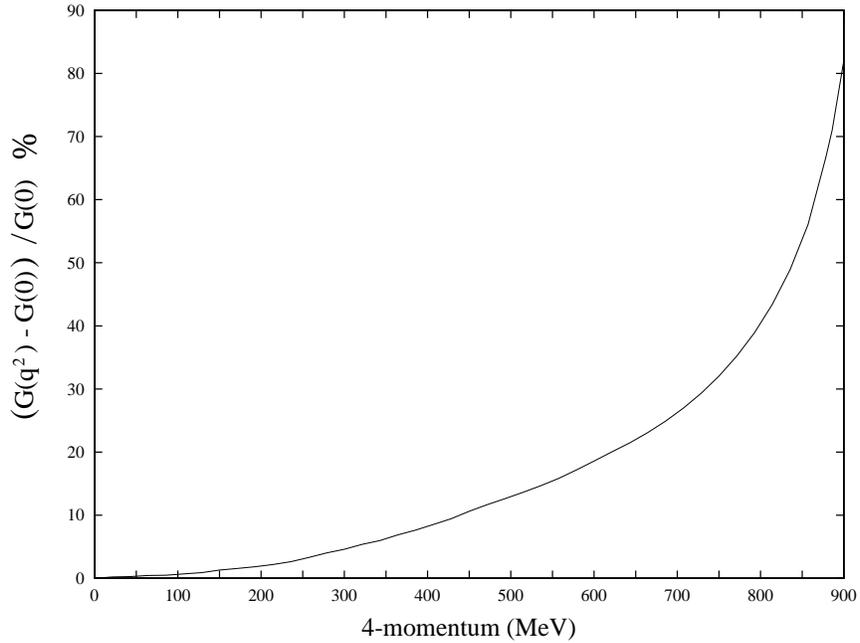,height=9.0cm,angle=-90}}
\protect\caption{ The 4-momentum dependence of the function 
$G(q^2, m_\pi^2)$, used in Eq. (\protect\ref{width}). 
Here the deviation from its value at $q^2 = 0$ is shown. }
\label{plot}
\end{figure}

\vspace*{2cm}

\begin{figure}
\centerline{\psfig{figure=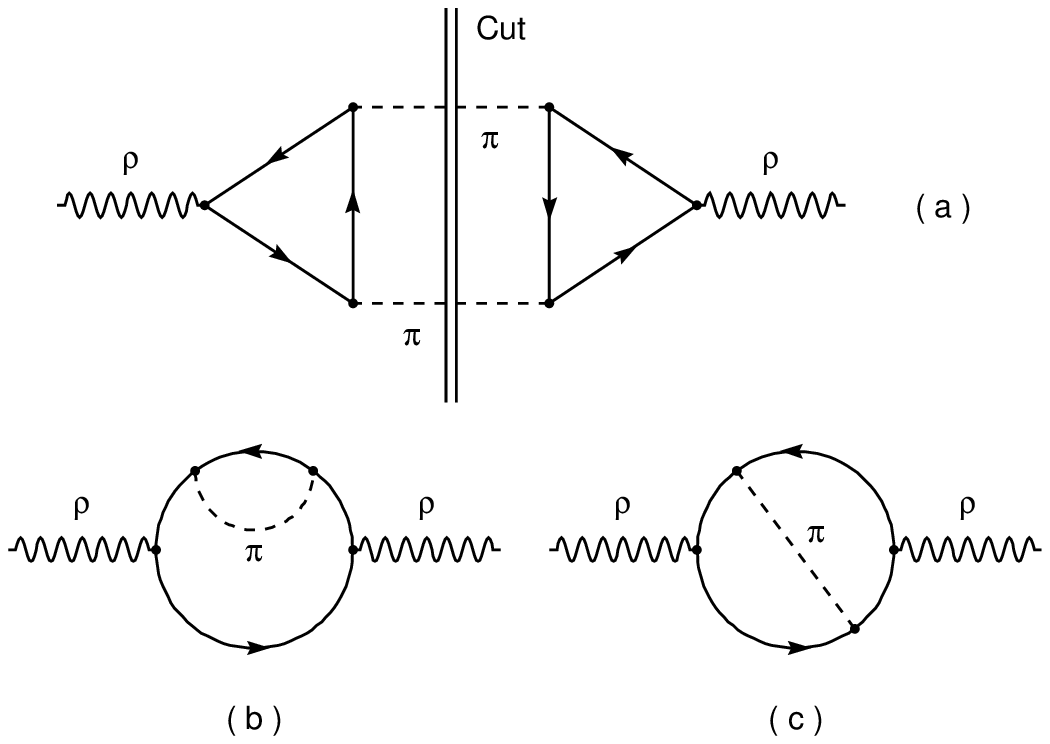,height=7.cm}} 
\protect\caption{Rho self-energy diagrams to order $1/N_c$. As
explained in the text, within our approximations only diagram (a), with
a cut across the pion lines, does contribute.}  
\label{bubbles}
\end{figure}


\begin{thebibliography}{25}

\bibitem{Sak69}
J.J. Sakurai, 
{\it Currents and mesons} (Univ. of Chicago Press, Chicago, 1969);
V. de Alfaro, S. Fubini, G. Furlan and C. Rossetti,
{\it Currents in Hadron Physics} (North Holland, Amsterdam, 1973). 

\bibitem{NJL61}
Y. Nambu and G. Jona-Lasinio,
Phys. Rev. {\bf 122}, 345 (1961);
Phys. Rev. {\bf 124}, 246 (1961).

\bibitem{VW91}
U. Vogl and W. Weise,
Prog. Part. Nucl. Phys. {\bf 27}, 195 (1991).

\bibitem{Kle92}
S. Klevansky,
Rev. Mod. Phys. {\bf 64}, 649 (1992).

\bibitem{HK94}
T. Hatsuda and T. Kunihiro,
Phys. Rep. {\bf 247}, 221 (1994).

\bibitem{Bij95}
J. Bijnens,  
Phys. Rep. {\bf 265}, 369 (1996).

\bibitem{BM88}
V. Bernard and U. G. Meissner,
Nucl. Phys. {\bf A489}, 647 (1988).

\bibitem{KLVW90}
S. Klimt, M. Lutz, U. Vogl and W. Weise,
Nucl. Phys. {\bf A516}, 429 (1990).

\bibitem{BHS88}
A. Blin, B. Hiller and M. Schaden,
Z. Phys. {\bf A331}, 75 (1988).

\bibitem{QK94}
E. Quack and S. Klevansky,
Phys. Rev. {\bf C49}, 3283 (1994).

\bibitem{Tho76}
G. t'Hooft,
Nucl. Phys. {\bf B75}, 461 (1974). 

\bibitem{Wit79}
E. Witten,
Nucl. Phys. {\bf B160}, 57 (1979).

\bibitem{BBDLS76}
D. R. Bes, R. A. Broglia, G. G. Dussel, R. J. Liotta and H. M. Sofia,
Nucl Phys. {\bf A260}, 1 (1976).

\bibitem{BBDLP76}
H. Reinhardt,
Nucl. Phys. {\bf A251}, 317 (1975);
D. Bes, R.A. Broglia, G.G. Dussel, R.J. Liotta and R.P.J. Perazzo,
Nucl. Phys. {\bf A260}, 77 (1976).

\bibitem{BE93KR91}
S. Krewald, K. Nakayama and J. Speth,
Phys. Lett. {\bf B272}, 190 (1991);
V. Bernard, A.H. Blin, B. Hiller, U-G. Mei\ss ner and M.C. Ruivo,
Phys. Lett. {\bf B385}, 163 (1993).

\bibitem{BMO94}
V. Bernard, U-G. Mei\ss ner and A.A. Osipov,
Phys. Lett. {\bf B324}, 201 (1994);
V. Bernard, A.H. Blin, B. Hiller, Y.P. Ivanov, A.A. Osipov and U-G. 
Mei\ss ner, Ann. Phys. (NY) {\bf 249}, 499 (1996).

\bibitem{MRBG94}
R. Alkofer, H. Reinhardt and H. Weigel,
Phys. Rep. {\bf 265}, 139 (1996);
Th. Meissner, E. Ruiz Arriola, A. Blotz and K. Goeke,
submitted to Rep. Prog. Phys.

\bibitem{DSTL95}
V. Dmitrasinovi\'c, H.J. Schulze, R. Tegen and H.R. Lemmer,
Ann. Phys. (NY) {\bf 238}, 332 (1995).

\bibitem{Dal61}
A. Dalgarno, 
in {\it Quantum Theory: I. Elements},
ed. D.R. Bates (Academic Press, New York, 1961), Ch.5.

\bibitem{PDT94}
Particle Data Group,
Phys. Rev. {\bf D50}, 1196 (1994).

\bibitem{BBB76}
P.F. Bortignon, R.A. Broglia and D.R. Bes,
Phys. Lett. {\bf B76} 153 (1976);
N.N. Scoccola and D.R. Bes,
Nucl. Phys. {\bf A425}, 256 (1984).

\bibitem{CSS92}
N-W. Cao, C.M. Shakin and W-D. Sun,
Phys. Rev. {\bf C46}, 2535 (1992).

\bibitem{KKW96}
F. Klingl, N. Kaiser and W. Weise, 
{\it hep-ph/9607431}.

\bibitem{MT96}
K. Mitchell and P. Tandy,
{\it nucl-th/9607025}. 

\bibitem{BB95}
R.D. Bowler and M.C. Birse,
Nucl. Phys. {\bf A582}, 655 (1995);
K. Langfeld and M. Rho,
Nucl. Phys. {\bf A596}, 451 (1996). 

\end{thebibliography}
\end{document}